\documentclass[sigconf]{acmart}
\AtBeginDocument{%
  \providecommand\BibTeX{{%
    \normalfont B\kern-0.5em{\scshape i\kern-0.25em b}\kern-0.8em\TeX}}}

\copyrightyear{2024}
\acmYear{2024}
\setcopyright{acmlicensed}\acmConference[WWW '24 Companion]{Companion
Proceedings of the ACM Web Conference 2024}{May 13--17, 2024}{Singapore,
Singapore}
\acmBooktitle{Companion Proceedings of the ACM Web Conference 2024 (WWW
'24 Companion), May 13--17, 2024, Singapore, Singapore}
\acmDOI{10.1145/3589335.3651472}
\acmISBN{979-8-4007-0172-6/24/05}

\usepackage{multirow}

\definecolor{oceanboatblue}{rgb}{0.0, 0.47, 0.75}

\definecolor{officegreen}{rgb}{0.0, 0.5, 0.0}

\newcommand{\ce}{CO$_2$E}

\definecolor{softgreen}{RGB}{0, 128, 0} 
\definecolor{softred}{RGB}{230, 62, 62} 
\definecolor{softyellow}{RGB}{162, 162, 30} 

\newcommand{\coesmall}[1]{\textbf{\textcolor{softgreen}{#1}}}
\newcommand{\coemedian}[1]{\textbf{\textcolor{softyellow}{#1}}}
\newcommand{\coelarge}[1]{\textbf{\textcolor{softred}{#1}}}

\begin{document}

\newcommand{\model}{GreenRec}

\title{Benchmarking News Recommendation in the Era of Green AI}

\author{Qijiong Liu}
\authornote{Both authors contributed equally to this research (co-first authors).}
\affiliation{%
  \institution{PolyU, Hong Kong}
  \country{}
}
\email{liu@qijiong.work}

\author{Jieming Zhu}
\authornotemark[1]
\affiliation{%
  \institution{Huawei Noah's Ark Lab, China}
  \country{}
}
\email{jiemingzhu@ieee.org}

\author{Quanyu Dai}
\affiliation{%
  \institution{Huawei Noah's Ark Lab, China}
  \country{}
}
\email{quanyu.dai@connect.polyu.hk}

\author{Xiao-Ming Wu}
\authornote{Corresponding author.}
\affiliation{%
  \institution{PolyU, Hong Kong}
  \country{}
}
\email{xiao-ming.wu@polyu.edu.hk}

\renewcommand{\shortauthors}{Qijiong Liu, Jieming Zhu, Quanyu Dai, \& Xiao-Ming Wu}

\definecolor{Clego}{RGB}{120,120,30}
\definecolor{Cmmenders}{RGB}{120,30,120}
\definecolor{Cgreen}{RGB}{30,120,30}
\begin{abstract}
    
Over recent years, news recommender systems have gained significant attention in both academia and industry, emphasizing the need for a standardized benchmark to evaluate and compare the performance of these systems. Concurrently, Green AI advocates for reducing the energy consumption and environmental impact of machine learning. To address these concerns, we introduce the first Green AI benchmarking framework for news recommendation, known as GreenRec\footnote{\textbf{\texttt{\textcolor{Cgreen}{GreenRec}}} has been published as part of the \textbf{  \texttt{\textcolor{Clego}{Lego}\textcolor{Cmmenders}{mmenders}}} benchmark, which is a modular framework for recommender systems. The source code and data for GreenRec can be found at \url{https://github.com/Jyonn/Legommenders}.}, and propose a metric for assessing the tradeoff between recommendation accuracy and efficiency. Our benchmark encompasses 30 base models and their variants, covering traditional end-to-end training paradigms as well as our proposed efficient only-encode-once (OLEO) paradigm. Through experiments consuming 2000 GPU hours, we observe that the OLEO paradigm achieves competitive accuracy compared to state-of-the-art end-to-end paradigms and delivers up to a 2992\% improvement in sustainability metrics.

\end{abstract}




\maketitle

\section{Introduction}

News recommender systems have become a crucial tool in the digital age, helping users discover relevant news articles and stay informed amidst the ever-increasing number of sources available.
To integrate news content knowledge into recommender systems, content encoders are employed to capture semantic news representations.
Various approaches have been explored in these efforts, including simple content encoders represented by CNN~\cite{kim-2014-convolutional} and Attention-based~\cite{vaswani2017attention} models~\cite{an2019neural,wu2019neural,wu2019nrms,li2022miner}, as well as emerging content encoders based on pretrained language models~\cite{zhang2021unbert,prec}.

Despite the success achieved by these studies, there is still a notable absence of standardized benchmarks and uniform evaluation protocols for news recommendation.
Consequently, even when common datasets like MIND~\cite{wu2020mind} are employed for evaluation, existing studies often use their own data partitions and apply unique preprocessing steps. The absence of uniformity in data preprocessing methods leads to non-reproducible and often conflicting experimental outcomes across different studies, and the use of non-standard data preprocessing techniques makes it challenging to compare and evaluate results between papers.


Meanwhile, the environmental impact has also begun to draw attention from the machine learning community, as training large-scale networks leads to a huge increase of carbon emission~\cite{spillo2023towards}. Therefore, a series of studies~\cite{wilson2020urban,przybyla2022using} explore ways to reduce energy consumption in line with Green AI principles~\cite{schwartz2020green}.

\begin{figure}[t]
  \centering
  \includegraphics[width=\linewidth]{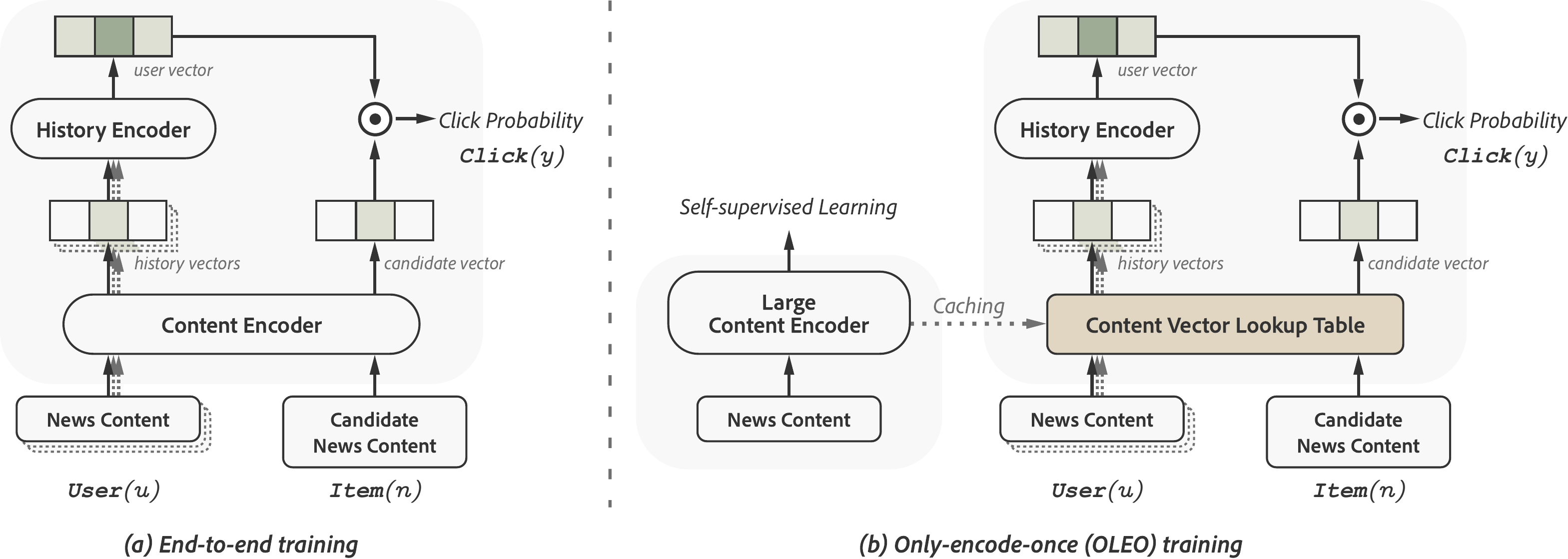}
  \caption{\label{fig:newsrec} 
  Two training paradigms for news recommendation.
  }
\end{figure}

In light of this, we establish the first Green AI Benchmark for News Recommendation, called \textbf{\model{}}, and introduce a sustainability metric, namely unit conversion rate of carbon emission, to strike a balance between recommendation quality and efficiency. Specifically, we  selected six base models and considered five different variants for each model. In total, we evaluated the recommendation accuracy and environmental impact across a total of 30 baselines.
This evaluation was carried out under two distinct training paradigms: the  \textit{\textbf{conventional end-to-end training paradigm}}, where the content encoder and other recommender components are jointly trained, and our  \textit{\textbf{proposed only-encode-once (OLEO) training paradigm}}, where the content encoder and other components are trained in a decoupled manner.
Our evaluation is conducted through hundreds of experiments within a standardized framework using the most widely used news recommendation datasets, MIND-small and MIND-large~\cite{wu2020mind}. 
The experiments reveal that the conventional end-to-end paradigm encounters efficiency challenges, while our OLEO-based paradigm is much more eco-friendly. Moreover, the OLEO-based model variants maintain competitive performance with the state-of-the-art end-to-end PLM-NR variants~\cite{wu2021empowering}, and achieve up to \textbf{2992\%} improvement in our sustainability metric.

\section{News Recommendation}

\subsection{Basics of News Recommender Systems }

News recommender systems are devised to predict the click probability of a user over a piece of given news article. Typically, a general news recommender system consists of three main components: content encoder, history encoder, and interaction module. 
As depicted in~\autoref{fig:newsrec} (a), the recommender accepts a news-user pair as input and yields the click probability. The content encoder is responsible for capturing the semantic meaning and contextual information of the news articles, while the history encoder processes the news history vectors, capturing the preferences, interests, and browsing patterns of users. Finally, the interaction module facilitates the interaction between the candidate news vector and user vector to calculate the click probability for the given news-user pair.

\subsection{Training Paradigms and Complexity}

Traditional models~\cite{wu2019npa,wu2019neural,wu2019nrms,wu2021empowering} typically follow such \textbf{end-to-end training paradigm}, as displayed in~\autoref{fig:newsrec} (a), where the three components are are trained together, guided by click-through data.

However, this traditional approach has shown limitations, particularly in terms of efficiency. A significant issue is the repetitive work done by the content encoder due to the repeated encoding of the same news articles, a consequence of their frequent interaction with various users. For instance, if there are $N$ news articles, $M$ users, and $K$ instances of user-news interactions, with an average historical sequence of $L$ interactions per user, the content encoder ends up processing $M \times L$ articles. The overall time required for training can be roughly calculated as $O(M \times L \times t_c + K \times (t_u + t_i))$, highlighting a particularly heavy load when $M \times L$ significantly exceeds either $N$ or $K$, especially with advanced content encoders~\cite{zhang2021unbert, wu2021empowering} that have a much higher time complexity compared to other modules.
For example, in the MIND dataset scenario, a single news article is encoded on average about \textcolor{softred}{\textbf{1818}} times in one epoch of training, indicating a substantial amount of redundancy.
To address these inefficiencies, we draw inspiration from a new training approach known as the \textbf{only-encode-once paradigm}, as depicted in~\autoref{fig:newsrec} (b). This approach involves initially pretraining the content encoder using the news content in a self-supervised manner. After this phase, the content vectors are extracted and stored, allowing the history encoder and interaction module to be trained separately with much quicker access to the news representations via a lookup table. The pretraining phase for the content encoder has a time complexity of $O(N \times t_c)$, considered a fixed cost across various models. Meanwhile, the training of the recommendation system itself becomes significantly more efficient, with a time complexity of $O(K \times (t_u + t_i))$, marking a drastic reduction from the traditional method.

\begin{table}[t]
\setlength\tabcolsep{2pt}
\resizebox{.9\linewidth}{!}{
\begin{tabular}{l|c|c|c|c|c|c|c}
\toprule
\small{Usage} & \small{TorchRec} & \small{DeepCTR} & \small{DeepRec} & \small{RecBole} & \small{FuxiCTR} & \small{BARS} & \small{Ours} \\
\midrule
\small{Data Preprocessing}    & \textcolor{softgreen}{\checkmark} & \textcolor{softgreen}{\checkmark} & \textcolor{softgreen}{\checkmark} & \textcolor{softgreen}{\checkmark} & \textcolor{softgreen}{\checkmark} & \textcolor{softgreen}{\checkmark} & \textcolor{softgreen}{\checkmark} \\
\small{Content-aware CTR}     & \coelarge{$\times$} & \coelarge{$\times$} & \coelarge{$\times$} & \coemedian{--}     & \coemedian{--}     & \coemedian{--}     & \textcolor{softgreen}{\checkmark} \\
\small{NewsRec Methods}       & \coelarge{$\times$} & \coelarge{$\times$} & \coelarge{$\times$} & \coelarge{$\times$} & \coelarge{$\times$} & \coelarge{$\times$} & \textcolor{softgreen}{\checkmark} \\
\small{Decoupled Training}    & \coelarge{$\times$} & \coelarge{$\times$} & \coelarge{$\times$} & \coemedian{--}     & \coemedian{--}     & \coemedian{--}     & \textcolor{softgreen}{\checkmark} \\
\small{End-to-end Training}   & \coelarge{$\times$} & \coelarge{$\times$} & \coelarge{$\times$} & \coelarge{$\times$} & \coelarge{$\times$} & \coelarge{$\times$} & \textcolor{softgreen}{\checkmark} \\
\small{Green AI Evaluation}   & \coelarge{$\times$} & \coelarge{$\times$} & \coelarge{$\times$} & \coelarge{$\times$} & \coelarge{$\times$} & \coelarge{$\times$} &  \textcolor{softgreen}{\checkmark} \\
\bottomrule
\end{tabular}
}
\caption{Comparison to exisiting recommendation benchmarks (\textcolor{softgreen}{\checkmark} | \coemedian{--} | \coelarge{$\times$} means totally | partially | not met, respectively). ``Partially met'' represents
incomplete availability.}\label{tab:benchmark-comparison}
\end{table}

\subsection{Adapting CTR Models to both Paradigms}

While the prevailing news recommendation models adhere to a matching-based approach with simple dot product as the interaction module, preserving all positive samples and conducting negative sampling (including real or fake negatives) and trained as a classification task, there are still some studies~\cite{zhang2021unbert,prec} that have validated the effectiveness of ranking-based CTR methods in news recommendation systems. In these approaches, all samples are retained throughout training, and the click probability is directly produced as a regression task. These CTR techniques incorporate intricate feature interaction networks, which can be utilized in the interaction module. When integrated into the content encoder, either through the end-to-end training paradigm or by initializing with pretrained news representations in the OLEO training paradigm, they gain the content-aware abilities.

\subsection{Proposed \model{} Benchmark}

Here, we summarize 6 representative methods and 5 variants used in our \model{} benchmark:

\textbf{Base models.} We select three representative news recommenders (i.e., NAML~\cite{wu2019neural}, LSTUR~\cite{an2019neural}, and NRMS~\cite{wu2019nrms}) and three widely used CTR methods (i.e., BST~\cite{bst}, DCN~\cite{wang2017deep}, and DIN~\cite{zhou2018deep})

\textbf{\coesmall{ID-based variants.}}\footnote{We use \coesmall{green}, \coemedian{yellow}, and \coelarge{red} text to represent \coesmall{OLEO-based or content-free (ID-based) training methods}, \coemedian{end-to-end training variants with simple content encoders} and \coelarge{end-to-end training variants with pretrained language models.}}
We remove the content encoder from the news recommenders and replace it with a \textbf{\textit{randomly initialized}} embedding lookup table. CTR methods are inherently ID-based.

\textbf{\coemedian{Text-based variants.}} We remove the news embedding lookup table of the CTR methods and replace it with a content sequence pooling layer. News recommenders are inherently text-based.

\textbf{\coelarge{PLM-NR variants.}} As proposed by PLM-NR~\cite{wu2021empowering}, we replace the content encoders of the text-based variants with pretrained language models such as BERT~\cite{devlin2018bert}.

\textbf{\coesmall{PREC variants.}} As proposed by PREC~\cite{prec}, we employ the OLEO training paradigm which first pretrains the content encoder and then replace the content encoders of the text-based variants with the \textbf{\textit{informative}} news embedding lookup table.

\textbf{\coesmall{BERT-OLEO variants.}} Unlike the PREC variants which requires the pretraining phase, here we directly use the pretrained BERT model to extract and cache the news representations, and construct the \textbf{\textit{informative}} lookup table.

\section{Implementation and Evaluation}

In this section, we first compare our \model{} with other benchmarking studies. Then, we outline our evaluation procedures, including quality-based assessments and our proposed sustainability metric. Finally, we report and discuss our benchmarking results. 

\subsection{Comparison with Previous Benchmarks}

In recent years, the issues of non-reproducibility and unfair comparisons have garnered increasing attention within the recommendation community. A series of studies~\cite{ivchenko2022torchrec,shen2017deepctr,zhang2019deeprec,zhao2022recbole,zhu2021open,zhu2022bars} have conducted comprehensive experiments and analyses, advocating for a unified benchmark construction. As shown in~\autoref{tab:benchmark-comparison}, we compare our news recommendation benchmark \model{} with these existing efforts. It is evident that these studies primarily target in the general recommendation domain, particularly emphasizing aspects such as CTR prediction and collaborative filtering. 
Nevertheless, a standardized benchmarking pipeline tailored to the news recommendation domain and the environmental evaluation remains conspicuously absent. Hence, our \model{} is the first attempt on benchmarking the news recommenders in the era of Green AI.

\subsection{Dataset and Data Splitting}

\begin{table}[t]
\centering
\renewcommand\arraystretch{1.2}

\resizebox{\linewidth}{!}{

\begin{tabular}{c|ccccc}
\toprule
 & $\#$ News & $\#$ Users & $\#$ Interactions & $\#$ Samples & Density \\
\midrule
\textbf{MIND-small} & 65,238 & 94,057 & 347,727 & 8,381,093 & 0.1366\% \\
\textbf{MIND-large} & 104,151 & 750,434 & 3,958,501 & 95,447,571 & 0.1221\% \\
\bottomrule
\end{tabular}

}
\caption{\label{tab:dataset} Dataset statistics. Density is defined as the ratio of \# interactions to \# all possible interactions.}
\end{table}

\begin{table*}[t]

\centering
\renewcommand\arraystretch{1.1}
\setlength\tabcolsep{5pt}

\resizebox{.85\linewidth}{!}{
\begin{tabular}{rl|ccc|ccc|ccc|ccc}
\toprule
\multicolumn{2}{c|}{\textbf{Dataset}}  & \multicolumn{6}{c|}{\textbf{MIND-small}} & \multicolumn{6}{c}{\textbf{MIND-large}} \\
\midrule
\multicolumn{2}{c|}{\multirow{2}{*}{\textbf{Method}}} & \multicolumn{3}{c|}{\textbf{Matching}} & \multicolumn{3}{c|}{\textbf{CTR}} & \multicolumn{3}{c|}{\textbf{Matching}} & \multicolumn{3}{c}{\textbf{CTR}} \\
& & \textbf{NAML} & \textbf{LSTUR} & \textbf{NRMS} & \textbf{BST} & \textbf{DCN} & \textbf{DIN} & \textbf{NAML} & \textbf{LSTUR} & \textbf{NRMS} & \textbf{BST} & \textbf{DCN} & \textbf{DIN} \\
\midrule
\multirow{5}{*}{\coesmall{ID-based}}
 & \textbf{AUC} & 50.13 & 51.04 & 54.84 & 50.09 
       & 53.92 & 55.95 
       & 52.98 & 54.98 & 57.59 & 52.10 
       & 57.41 & 57.36 \\
 & \textbf{MRR} & 23.01 & 22.90 & 26.53 & 22.13 
       & 25.18 & 25.88 
       & 24.52 & 25.99 & 27.41 & 24.81 
       & 26.76 & 26.70 \\
 & \textbf{N@5} & 22.35 & 22.31 & 26.34 & 21.59 
       & 24.43 & 25.95 
       & 24.12 & 25.64 & 27.05 & 24.63 
       & 26.90 & 26.84 \\
 & \textbf{\ce $\downarrow$} & \textbf{19}    & \textbf{20}    & \textbf{28}    & \textbf{38}    
       & \textbf{60}    & \textbf{84}    
       & \textbf{294}   & \textbf{353}   & \textbf{471}   & \textbf{555}   
       & \textbf{926}   & \textbf{1294}   \\
 & \textbf{ApC} & 0.68  & 5.20  & 17.29 & 0.24  
       & 6.53  & 7.08
       & 1.01  & 1.41  & 1.61  & 0.38  
       & 0.80  & 0.57  \\
\midrule
\multirow{5}{*}{\coemedian{Text-based (End-to-end)}} 
 & \textbf{AUC} & 60.14 & 61.27 & 62.21 & 60.51 
       & 62.63 & 62.90 
       & 63.03 & 63.89 & 64.12 & 63.28 
       & 63.88 & 64.02 \\
 & \textbf{MRR} & 28.93 & 29.64 & 30.19 & 28.59 
       & 29.73 & 30.06 
       & 30.40 & 31.24 & 31.77 & 30.73 
       & 31.65 & 31.98 \\
 & \textbf{N@5} & 29.33 & 30.28 & 31.10 & 29.09 
       & 30.52 & 30.65 
       & 31.82 & 32.15 & 32.64 & 31.95 
       & 32.40 & 33.00 \\
 & \textbf{\ce $\downarrow$} & 42    & 58    & 62    & 53    
       & 63    & 90    
       & 648   & 892   & 1010  & 1212  
       & 972  & 1386  \\
 & \textbf{ApC} & 24.14 & 19.43 & 19.69 & 19.83 
       & 20.05 & 14.33 
       & 2.01  & 1.56  & 1.40  & 1.09  
       & 1.43  & 1.01  \\
\midrule
\multirow{5}{*}{\coelarge{PLM-NR (End-to-end)}}
 & \textbf{AUC} & \underline{62.06} & \textbf{63.64} & \underline{62.53} & \textbf{64.40} 
       & \underline{63.32} & \textbf{63.26} 
       & \textbf{65.19} & \textbf{65.73} & \textbf{65.57} & \textbf{66.03} 
       & \underline{65.42} & \textbf{65.31} \\
 & \textbf{MRR} & \textbf{31.66} & \textbf{31.74} & \underline{30.74} & \textbf{32.21} 
       & \underline{32.00} & \textbf{31.83} 
       & \textbf{32.74} & \textbf{33.18} & \textbf{32.94} & \textbf{33.40} 
       & \underline{32.85} & \underline{32.68} \\
 & \textbf{N@5} & \textbf{32.25} & \textbf{32.72} & \underline{31.31} & \textbf{33.34} 
       & \underline{32.58} & \textbf{32.40} 
       & \textbf{33.77} & \textbf{34.26} & \textbf{34.13} & \textbf{34.70} 
       & \underline{33.99} & \textbf{33.70} \\
 & \textbf{\ce $\downarrow$} & 178   & 202   & 252   & 505   
       & 1,752   & 1,839   
       & 2,527  & 3,032  & 4,043  & 8,086  
       & 27,036  & 28,329 \\
 & \textbf{ApC} & 6.78  & 6.75  & 4.97  & 2.85  
       & 0.76  & 0.72  
       & 0.60  & 0.52  & 0.39  & 0.20  
       & 0.06  & 0.05  \\
\midrule
\multirow{5}{*}{\coesmall{BERT (OLEO)}}
 & \textbf{AUC} & 60.62 & 61.09 & 60.94 & 60.81 
       & 62.65 & 62.40 
       & 63.02 & 63.62 & 63.40 & 62.94 
       & 64.29 & 63.75 \\
 & \textbf{MRR} & 29.31 & 29.26 & 29.31 & 29.04 
       & 30.92 & 30.75 
       & 31.23 & 31.59 & 31.38 & 30.56 
       & 32.60 & 31.58 \\
 & \textbf{N@5} & 29.71 & 29.60 & 29.65 & 29.38 
       & 31.37 & 32.44 
       & 31.79 & 32.30 & 32.16 & 31.83 
       & 33.63 & 32.43 \\
 & \textbf{\ce $\downarrow$} & \underline{22}    & \underline{23}    & \underline{33}    & \underline{38}    
       & \underline{62}    & \underline{86}    
       & \underline{353}   & \underline{404}   & \underline{505}   & \underline{640}   
       & \underline{956}  & \underline{956}   \\
 & \textbf{ApC} & \underline{48.27} & \underline{48.22} & \underline{33.15} & \underline{28.45} 
       & \underline{20.40} & \underline{14.41} 
       & \underline{3.69}  & \underline{3.37}  & \underline{2.65}  & \underline{2.02}  
       & \underline{1.49}  & \underline{1.44}   \\
\midrule
\multirow{5}{*}{\coesmall{PREC (OLEO)}}
 & \textbf{AUC} & \textbf{62.95} & \underline{62.16} & \textbf{62.95} & \underline{62.43} 
       & \textbf{64.57} & \underline{63.12} 
       & \underline{64.78} & \underline{64.88} & \underline{64.34} & \underline{65.33} 
       & \textbf{65.44} & \underline{64.53} \\
 & \textbf{MRR} & \underline{31.26} & \underline{31.00} & \textbf{31.18} & \underline{30.42} 
       & \textbf{32.60} & \underline{31.28} 
       & \underline{32.64} & \underline{32.94} & \underline{32.93} & \underline{33.29} 
       & \textbf{33.04} & \textbf{32.72} \\
 & \textbf{N@5} & \underline{32.01} & \underline{31.79} & \textbf{32.10} & \underline{30.94} 
       & \underline{33.48} & \underline{32.01} 
       & \underline{33.66} & \underline{34.00} & \underline{33.95} & \underline{34.35} 
       & \textbf{34.03} & \underline{33.58} \\
 & \textbf{\ce $\downarrow$} & \underline{22}    & \underline{23}    & \underline{33}    & \underline{38}    
       & \underline{62}    & \underline{86}    
       & \underline{353}   & \underline{404}   & \underline{505}   & \underline{640}   
       & \underline{956}  & \underline{956}   \\
& \textbf{ApC} & \textbf{58.86} & \textbf{52.87} & \textbf{39.24} & \textbf{32.71} 
       & \textbf{23.50} & \textbf{15.25} 
       & \textbf{4.19}  & \textbf{3.68}  & \textbf{2.84}  & \textbf{2.40}  
       & \textbf{1.61}  & \textbf{1.52} \\ \midrule
\multicolumn{2}{r|}{\textbf{ApC Imp. (\%)}} & 768\% & 683\% & 690\% & 1048\% & 2992\% & 2018\% & 598\% & 608\% & 628\% & 1100\% & 2583\% & 2940\% \\
\bottomrule
\end{tabular}
}

\caption{Comparison of variants of different recommenders. ``ApC Imp'' represents the environmental sustainability growth rate of PREC variants compared to the SOTA PLM-NR variants. We bold the best results and underline the second-best results.
}
\label{tab:big-table}

\end{table*}


In this work, we mainly use two versions of the most prevalent news recommendation dataset: MIND-small and MIND-large~\cite{wu2020mind}. \autoref{tab:dataset} summarizes the data statistics information. 

Due to the fact that the MIND-small dataset only provides training and validation sets, many studies have failed to specify their exact partitioning during training, and in some cases, they have directly reported results on the validation set, which leads to an unfair comparison among different models. Furthermore, the MIND-large dataset only offers online testing, which is not conducive to offline environment assessment. To address these issues, we have restructured both datasets by splitting the validation sets of both datasets in a 1:1 ratio to create new validation and testing sets. The scripts can be found in our released repository.

\subsection{Evaluation Metrics}

We employ three widely-used metrics for accuracy evaluation, including AUC, MRR, nDCG@5~\cite{wu2020mind}.
Moreover, we also introduce one Green AI metrics: Carbon Emission~\cite{lacoste2019quantifying} (\ce) and AUC per Carbon Emission (ApC), to evaluate the sustainability and energy utilization of model $\Psi$. $CO_2E$ is computed by:
\begin{equation}
    CO_2E = p \times t \times c,
\end{equation}
where $p$ is the power consumption differ by hardware types, $t$ is the hardware running time, and $c$ is carbon efficiency differ by power provider. ApC measures the unit conversion rate of carbon emissions on AUC, defined as:
\begin{equation}
    ApC = \frac{AUC - 50}{CO_2E} \times 100.
\end{equation}
\textbf{Training Details.} The Adam optimizer is employed for optimization. We consider various learning rates from the set $\{1e-5, 2e-5, 5e-5, 1e-4, 2e-4, 5e-4\}$ and batch sizes from the set \\ $\{64, 128, 256, 500, 1000, 5000\}$. The embedding dimensions for all base models and their variants are fixed to 64. For the OLEO training paradigm which uses cached news vectors for initialization, we keep the pretrained embeddings fixed and use learnable transformation matrices for dimension projection. Experimental results are averaged over five runs, and all methods were trained using Nvidia GeForce RTX 3090 with 24GB memory, which has a power consumption of 350W. The carbon efficiency in our experimental region (anonymous for review) is 722g CO2-eq/kWh.

\subsection{Findings}

The comparative results of various models are summarized in \autoref{tab:big-table}, from which we derive the following observations.
    
\textbf{ID- vs. Modality-based Recommender.} Among the five variants, the ID-based variants exhibit the worst performance, showing the significance of news content comprehension in news recommenders. The findings in recent study~\cite{yuan2023go} are in line with ours.

\textbf{Enhancement of PLMs in the end-to-end training paradigm.} The PLM-NR variants consistently outperforms the ID-based and Text-based ones, benefiting from the augmented understanding of news content offered by pretrained knowledge. However, they grapple with an efficiency issue, resulting in significantly higher carbon emissions compared to other variants.

\textbf{Enhancement of PLMs in the OLEO training paradigm.} The BERT variants maintain an environmentally sustainable level of efficiency. Nonetheless, there is a deficiency in the ability of general language models to fully grasp news content, resulting in a notable decline in accuracy compared to the PLM-NR variants. However, PREC variants achieve comparable performance to PLM-NR ones while significantly reducing carbon emissions, which represents the prime case of the tradeoff between accuracy and efficiency. 

\section{Conclusion}

We have established the first standardized Green AI benchmark for news recommendation. Through extensive experiments over 30 base models and their variants, we conclude that the OLEO training paradigm successfully strikes a balance between accuracy and efficiency. We encourage other researchers to explore the OLEO paradigm and make contributions to environmental conservation.

\section*{Acknowledgement}

Qijiong Liu would like to thank Dr. Guanliang Chen and Mr. Haoxiang Shi for their insightful input during the brainstorming process of our paper. Additionally, the authors would like to thank the anonymous reviewers for their valuable feedback.

\bibliographystyle{ACM-Reference-Format}
\bibliography{GreenRec}

\end{document}